\documentclass[aps,draft]{revtex4}
\usepackage[dvips]{graphicx}
\begin{document}

\title{Excitation of Longitudinal Waves in a Degenerate Isotropic Quantum Plasma }
\author{ Levan N.Tsintsadze }
\thanks{Also at Department of Plasma Physics, E.Andronikashvii Institute of Physics, Tbilisi,
Georgia}
\affiliation{Graduate School of Science, Hiroshima University, Higashi-Hiroshima, Japan}
\author{ Nodar L.Tsintsadze }
\affiliation{Department of Plasma Physics, E.Andronikashvili
Institute of Physics, Tbilisi, Georgia}

\date{\today}

\begin{abstract}

A dispersion equation, which describes the interaction of low density electron beam with a degenerate electron quantum plasma, is derived and examined for some interesting cases. In addition to the instabilities similar to those for classical plasma, due to the quantum effect a new type of instability is found. Growth rates of these new modes, which are purely quantum, are obtained. Furthermore, the excitation of Bogolyubov's type of spectrum by a strong electric field is discussed.

\end{abstract}


\maketitle
Quantum plasmas are common in planetary interiors, in compact astrophysical objects, in conductors, semiconductors and micro-mechanical systems, as well as in the next generation intense laser-solid density plasma experiments. The field of quantum plasma physics is evolving, though it has a long and diverse tradition there are many questions and issues which one has to address.
In the past Klimontovich and Silin \cite{kli} have studied the properties of linear electron plasma oscillations in a dense Fermi plasma, and since then a huge number of works have been published on collective behavior of quantum plasmas using a set of incorrect hydrodynamic equations \cite{manf}, \cite{mar}. In the previous paper \cite{tsin} this problem has been solved by deriving a new type of quantum kinetic equations of the Fermi particles of various species, and a general
set of fluids equations describing the quantum plasma was obtained. This novel kinetic equation for the Fermi quantum plasma was used in Ref.\cite{tsin} to study the propagation of small longitudinal perturbations in an electron-ion collisionless plasmas, deriving a quantum dispersion equation. The effects of the quantization of the orbital motion of electrons and the spin of electrons on the propagation of longitudinal waves in the quantum plasma has been also reported very recently \cite{ltsin}.

In the present paper, we study an electron beam - plasma instabilities and excitation of Bogolyubov's type of spectrum in the Fermi quantum plasma. For our purpose, we  employ the quantum kinetic equation derived in Ref.\cite{tsin}, which reads
\begin{eqnarray}
\label{tob}
\frac{\partial f_\alpha }{\partial t}+\left( \vec{v}\cdot\nabla \right) f_\alpha
+e_\alpha \Bigl(\vec{E}+\frac{\vec{v}_\alpha \times \vec{H}}{c}\Bigr)\frac{\partial f_\alpha }{
\partial \vec{p}}+\frac{\hbar^2}{2m_\alpha }\nabla \frac{1}{\sqrt{
n_\alpha }}\Delta \sqrt{n_\alpha }\ \frac{\partial f_\alpha }{\partial \vec{p}}=C(f_\alpha )\ ,
\end{eqnarray}
where suffix $\alpha$ stands for the particle species, $\hbar$ is the Planck constant divided by $2\pi$, $C(f_\alpha )$ is the collision integral and the other notation is standard. It should be emphasized that this equation elaborates all the information on
the quantum effects. We specifically note also that this equation is rather simple from the mathematically point of
view.

Considering the propagation of small longitudinal
perturbations ($\vec{H}=0$, $\vec{E}=-\nabla \varphi $) in an electron-ion collisionless plasmas
for a weak field, we look for the electron and ion distribution functions in the form $f_\alpha =f_{\alpha 0}+\delta f_\alpha $, where $f_{\alpha 0}$ is the stationary isotropic homogeneous distribution function unperturbed by the field, and $\delta f_\alpha $ is the small variation in it due to the field.
We linearize Eq.(\ref{tob}) with respect to the perturbation and assume $\delta f_\alpha $ and $\delta \varphi $ vary like $ \exp{i(\vec{k}\cdot\vec{r}-\omega t)}$.

Using the Poisson's equation
\begin{eqnarray}
\label{puas}
\Delta \delta \varphi =4\pi e\left\{ 2\int \frac{d^3p}{(2\pi \hbar )^3}
\delta f_e-2\int \frac{d^3p}{(2\pi \hbar )^3}\delta f_i\right\}
\end{eqnarray}
and assuming the Fermi degeneracy temperature $T_{F}=\frac{\varepsilon_F}{K_B}$ ($K_B$ is the Boltzmann coefficient, the Fermi distribution function is the step function $f_{\alpha 0}=\Theta (\varepsilon_{F\alpha }-\varepsilon )$, where
$\varepsilon_{F\alpha }=\frac{m_\alpha v_{F\alpha }^2}{2}\ $) much higher than the Fermi gas temperature,
we then obtain after some algebra the quantum dispersion equation \cite{tsin}
\begin{eqnarray}
\label{qde}
\varepsilon =1+\sum_\alpha\frac{3\omega_{p\alpha}^2}{\Gamma_\alpha k^2v_{F\alpha}^2}\left\{1-\frac{\omega }{2kv_{F\alpha}}\ln
\frac{\omega +kv_{F\alpha}}{\omega -kv_{F\alpha}}\right\}=0 \ ,
\end{eqnarray}
where
\[
\Gamma_\alpha=1+\frac{3\hbar^2k^2}{4m_\alpha v_{F\alpha}^2}\Bigl(1-\frac{\omega }{
2kv_{F\alpha}}\ln \frac{\omega +kv_{F\alpha}}{\omega -kv_{F\alpha}}\Bigr)\ ,
\]
and $\omega$ can be more or less than $kv_{Fe}$. Note that for $\omega \gg kv_{Fi},\ $ $\Gamma_i\approx 1.$

For the electron Langmuir waves supposing the ion
mass $m_i\rightarrow \infty $ and $\omega \gg kv_{Fe},\ $ or the
range of fast waves, when the phase velocity exceeds the Fermi velocity of
electrons, we recover the dispersion relation derived by Klimontovich and Silin \cite{kli}
\begin{eqnarray}
\label{dlw}
\omega^2=\omega_{pe}^2+\frac{3k^2v_{Fe}^2}{5}+\frac{\hbar^2k^4}{4m_e^2}+...
\end{eqnarray}
where $\omega_{pe}$ is the electron plasma frequency.
This expression shows that the high frequency oscillations of electrons of a degenerate
plasma remain undamped in the absence of particles collisions. Note that the
Landau damping is also absent, since according to the Fermi distribution there
are no particles with velocities greater than the Fermi velocity which could
contribute to the absorption.

We now propose the excitation of the Klimontovich and Silin's spectrum (\ref{dlw}) by a straight electron beam with the density $n_b$ much less than the plasma density, which is injected into a degenerate electron gas. The electron beam is assumed to be classical (since the density is low) obeying the Maxwellian distribution function with non-relativistic temperatures.
Taking into account the Madelung term in the kinetic equation of the beam electrons, we write down the desired dispersion equation for the plasma - beam system
\begin{eqnarray}
\label{dde}
1+\delta\varepsilon_p+\delta\varepsilon_b=0
\end{eqnarray}
\begin{eqnarray*}
1+\frac{3\omega_{pe}^2}{k^2v_{Fe}^2\Gamma_e}\left\{1-\frac{\omega}{2kv_{Fe}}\ln\frac{\omega+kv_{Fe}}{\omega-kv_{Fe}}\right\}+
\frac{\omega_b^2}{k^2v_{tb}^2\Gamma_b}\left\{1-I_+\Bigl(\frac{\omega-\vec{k}\cdot\vec{u}}{kv_{tb}}\Bigr)\right\}=0 \ ,
\end{eqnarray*}
where $\vec{u}$ is the velocity of beam, $v_{tb}=\sqrt{T/m_e}$ and $\omega_b$ are the thermal speed and the Langmuir frequency of the beam electrons, respectively, the function $I_+(x)=xe^{-x^2/2}\int_{i\infty}^xd\tau e^{\tau^2/2}$ has been studied in details in Ref.\cite{ale} and has such asymptotes
\begin{eqnarray}
\label{as1}
I_+(x)=1+\frac{1}{x^2}+\frac{3}{x^4}+...-i\sqrt{\frac{\pi}{2}}xe^{-x^2/2} \hspace{.7cm} if \hspace{.7cm} x\gg 1
\end{eqnarray}
\begin{eqnarray}
\label{as2}
I_+(x)\simeq -i\sqrt{\frac{\pi}{2}}x \hspace{1cm} if \hspace{1cm} x\ll 1 \ .
\end{eqnarray}
With the asymptote (\ref{as1}) at hand, supposing $\omega\gg kv_{Fe}$ we obtain
\begin{eqnarray}
\label{sdi}
1-\frac{\omega_L^2}{\omega^2}-\frac{\omega_b^2(1+\iota\beta)}{(\omega-\vec{k}\cdot\vec{u})^2-\omega_q^2(1+\iota\beta)}=0 \ ,
\end{eqnarray}
where $\omega_L$ is the Klimontovich-Silin dispersion relation (\ref{dlw}), $\omega_q=\hbar k^2/2m_e$ is the frequency of quantum oscillations of electron, and $\beta$ is due to Landau damping
\begin{eqnarray}
\label{beta}
\beta=-\sqrt{\frac{\pi}{2}}\frac{(\omega-\vec{k}\cdot\vec{u})^3}{(kv_{tb})^3}\exp\left\{-\frac{(\omega-\vec{k}\cdot\vec{u})^2}
{2k^2v_{tb}^2}\right\} \ .
\end{eqnarray}

We examine the dispersion equation (\ref{sdi}), which describes the interaction of low density electron beam with the degenerate electron plasma, for some interesting cases. First, we assume that the electron beam is a monoenergetic straight beam, so that the thermal motion of electrons of the beam is neglected, $\beta=0$. In this case Eq.(\ref{sdi}) reduces to the following dispersion relation, which was recently discussed by Kuzelev and Rukhadze \cite{kuz}
\begin{eqnarray}
\label{fcd}
1-\frac{\omega_L^2}{\omega^2}-\frac{\omega_b^2}{(\omega-\vec{k}\cdot\vec{u})^2-\omega_q^2}=0 \ .
\end{eqnarray}

As is known in the classical case ($\omega_q=0$) the interaction of an electron beam with a plasma is strong when the Cherenkov resonance condition ($\omega\simeq\vec{k}\cdot\vec{u}$) is fulfilled. However, in the quantum plasma $\omega_q\neq 0$, and the quantum interaction is governed by first-order poles (the last term in Eq.(\ref{fcd})). Hence, the quantum frequency $\omega_q$ of the de Broglie waves in Eq.(\ref{fcd}) leads to the Doppler resonance alone, $\omega=\vec{k}\cdot\vec{u}-\omega_q$. Equation (\ref{fcd}) therefore admits the solution at $\omega=\omega_L+\gamma$ and $\omega=\vec{k}\cdot\vec{u}-\omega_q+\gamma$, with $\mid\gamma\mid\ll\omega$
\begin{eqnarray}
\label{dsol}
Im\gamma=\frac{\omega_b}{2}\Bigl(\frac{\omega_L}{\omega_q}\Bigr)^{1/2}\ .
\end{eqnarray}
This expression describes the excitation of the Klimontovich-Silin waves by the monoenergetic electron beam. It should be noted that the growth rate is purely quantum.

Next, we consider the kinetic instability. Noting that $\mid\beta\mid\ll 1$, in Eq.(\ref{sdi}) we neglect $\beta$ in the denominator at $(\omega-\vec{k}\cdot\vec{u})^2\neq\omega_q^2$  and the beam contribution to the real part in the numerator. With this assumption the dispersion relation (\ref{sdi}) casts in to the form
\begin{eqnarray}
\label{kins}
1-\frac{\omega_L^2}{\omega^2}-\frac{\iota\omega_b^2\beta}{(\omega-\vec{k}\cdot\vec{u})^2-\omega_q^2}=0 \ ,
\end{eqnarray}
which has a solution of the form $\omega=\omega_L+\iota\gamma$, and the growth rate is
\begin{eqnarray}
\label{kgr}
\gamma=-\sqrt{\frac{\pi}{8}}\frac{\omega_L\omega_b^2}{(kv_{tb})^3}\frac{
(\omega_L-\vec{k}\cdot\vec{u})^3}{(\omega_L-\vec{k}\cdot\vec{u})^2-\omega_q^2}\exp\left\{-\frac{(\omega_L-\vec{k}\cdot\vec{u})^2}
{2k^2v_{tb}^2}\right\} \ .
\end{eqnarray}
We note here that at $\vec{k}\cdot\vec{u}>\omega_L$, $\ \gamma >0$ and the growth rate due to the Cherenkov resonance is larger than the classical one at $\omega_q=0$.

Equation (\ref{sdi}) has an another solution due to the quantum effect. Namely, for the range of frequencies $\omega<\omega_L,\ \vec{k}\cdot\vec{u}$ and $\vec{k}\cdot\vec{u}\simeq\omega_q$, in which case $\beta$ is positive, from Eq.(\ref{sdi}) we get
\begin{eqnarray}
\label{ann}
\frac{\omega_L^2}{\omega^2}+\iota\frac{\omega_b^2}{\omega_q^2\beta}=0 \ .
\end{eqnarray}
The solution of which is
\begin{eqnarray}
\label{sann}
\omega=\frac{1+\iota}{\sqrt{2}}\ \frac{\omega_L\omega_q}{\omega_b}\ \beta^{1/2} \ .
\end{eqnarray}
It should be emphasized that this instability has no analogy in the classical plasma.

We now return to Eq.(\ref{qde}) and introduce the Thomas-Fermi screening wave vector $k_{TF}=\frac{\sqrt{3}\omega_{pe}}{v_{Fe}}.\ $ Noting that in the limit $k^2\gg k_{TF}^2,\ $ $\omega $ tends to $kv_{Fe}\ $ (at $\ m_i\rightarrow \infty )\ $ we obtain from Eq.(\ref{qde})
\begin{eqnarray}
\label{klim}
\omega =kv_{Fe}\Bigl(1+2\exp \{-\frac{2(\frac{k^2}{k_{TF}^2}+
\frac{\omega_q^2}{\omega_{pe}^2})}{1+\frac{\omega_q^2}{\omega_{pe}^2}}\}\Bigr)\ .
\end{eqnarray}
If we neglect the quantum term $\omega_q$ in Eq.(\ref{klim}), then
we recover waves known as the zero sound, which are the continuation of the
electron Langmuir wave (\ref{dlw}) into the range of short wavelength. Thus the
expression (\ref{klim}) represents the quantum correction to the zero sound.

In the following we discuss the excitation of the zero sound by a low density cold beam. As we have shown above at $\omega=\vec{k}\cdot\vec{u}-\omega_q+\gamma$, $\ \mid\gamma\mid\ll\omega$ the dielectric permittivity of the beam is
\begin{eqnarray}
\label{bedp}
\delta\varepsilon_b=\frac{\omega_b^2}{2\omega_q\gamma} \ .
\end{eqnarray}
Whereas in the dielectric permittivity of the plasma $\delta\varepsilon_p$ we suppose that
\begin{eqnarray}
\label{sup}
\omega\approx kv_{Fe}+\gamma \ .
\end{eqnarray}
Substituting Eq.(\ref{bedp}) into Eq.(\ref{dde}) and taking into account (\ref{sup}), we get the spectrum (\ref{klim}) for real $\omega$ and for the growth rate
\begin{eqnarray}
\label{zgr}
Im\gamma=\frac{\omega_Lkv_{Fe}}{\omega_L^2+\omega_q^2}\ \omega_b
\Bigl(\frac{2kv_{Fe}}{\omega_q}\exp \{-\frac{2(\frac{k^2}{k_{TF}^2}+
\frac{\omega_q^2}{\omega_L^2})}{1+\frac{\omega_q^2}{\omega_L^2}}\}\Bigr)\ .
\end{eqnarray}
Note that this growth rate is also purely quantum.

We next consider the excitation of Bogolyubov's type of spectrum, derived in the previous paper \cite{tsin}, by a strong electric field. We recall here that in the adiabatic approximation the velocity of electron may be regarded as constant (ions are assumed to be immobile). The electrons part of the dielectric permittivity now reads
\begin{eqnarray}
\label{epdp}
\delta\varepsilon_e=\frac{3\omega_{pe}^2}{k^2v_{Fe}^2}\ \frac{1}{\Gamma_e}\ \left\{1-\frac{\omega-\vec{k}\cdot\vec{u}}{2kv_{Fe}}\ln\frac{\omega-\vec{k}\cdot\vec{u}+kv_{Fe}}{\omega-\vec{k}\cdot\vec{u}-
kv_{Fe}}\right\} \ .
\end{eqnarray}
Here
\begin{eqnarray*}
\Gamma_e=1+\frac{3\omega_q^2}{k^2v_{Fe}^2}\left\{1-\frac{\omega-\vec{k}\cdot\vec{u}}{2kv_{Fe}}\ln\frac{
\omega-\vec{k}\cdot\vec{u}+kv_{Fe}}{\omega-\vec{k}\cdot\vec{u}-kv_{Fe}}\right\} \ .
\end{eqnarray*}
In the range of frequencies $\omega\gg kv_{Fi}$ and $\mid\omega-\vec{k}\cdot\vec{u}\mid\ll kv_{Fe}$ from Eq.(\ref{qde}), where $\omega$ is displaced by $\omega-\vec{k}\cdot\vec{u}$, we obtain
\begin{eqnarray}
\label{nedp}
\delta\varepsilon_e=\frac{\frac{3\omega_{pe}^2}{k^2v_{Fe}^2}\Bigl(1+\iota\sqrt{\frac{\pi}{2}}\ \frac{\omega-\vec{k}\cdot\vec{u}}{
kv_{Fe}}\Bigr)}{1+\frac{3\omega_q^2}{k^2v_{Fe}^2}\Bigl(1+\iota\sqrt{\frac{\pi}{2}}\ \frac{\omega-\vec{k}\cdot\vec{u}}{
kv_{Fe}}\Bigr)}
\end{eqnarray}
and
\begin{eqnarray}
\label{iedp}
\delta\varepsilon_i=-\frac{\omega_{pi}^2}{\omega^2} \ .
\end{eqnarray}
One can immediately see that Eq.(\ref{qde}) reduces to the simple dispersion relation $\delta\varepsilon_e+\delta\varepsilon_i=0$, solutions ($\omega=\omega^\prime+\iota\omega^{\prime\prime}$) of which are the Bogolyubov's type of spectrum \cite{tsin}
\begin{eqnarray}
\label{bog}
\omega^\prime =k\sqrt{\frac{2\varepsilon_{Fe}}{3m_i}+\frac{\hbar^2k^2}{4m_i m_e}}
\end{eqnarray}
and the imaginary part
\begin{eqnarray}
\label{bgrg}
\omega^{\prime\prime}=-\frac{\pi }{12}\ \frac{kp_{Fe}}{m_i}\ \frac{\omega^\prime-\vec{k}\cdot\vec{u}}{\omega^\prime} \ .
\end{eqnarray}
For an instability the following inequality should be satisfied
\begin{eqnarray}
\label{sin}
u>\frac{1}{\cos\theta}\sqrt{\frac{p_F^2}{3m_e m_i}+\frac{\hbar^2k^2}{4m_e m_i}} \ .
\end{eqnarray}
In the case of absence of the external field ($\vec{u}=0$) Eq.(\ref{bgrg}) reduces to the expression of damping rate obtained in Ref.\cite{tsin}
\begin{eqnarray}
\label{im}
\omega^{\prime\prime}=-\frac{\pi }{12}k\ \frac{p_{Fe}}{m_i} \ ,
\end{eqnarray}
which is much less than $\omega^\prime .$ The expression (\ref{im}) indicates that the damping rate
is determined by the electrons alone.

To summarize, we have investigated the excitation of longitudinal waves in a degenerate non-magnetized quantum plasma. Namely, we derived the dispersion equation, which describes the interaction of low density electron beam with the degenerate electron quantum plasma, and examined it for some interesting cases. In particular, we have shown the excitation of Klimontovich-Silin waves by the monoenergetic electron beam, the growth rate of which is purely quantum. Next, we studied the kinetic instability demonstrating that the growth rate due to the Cherenkov resonance is larger than the classical one. In addition, we disclosed a novel instability due to the quantum effect, which has no analogy in the classical plasma. We obtained the growth rate of this new mode. The excitation of the zero sound by a low density cold beam is also discussed and the growth rate, which is also purely quantum, is derived. Furthermore, we considered the excitation of Bogolyubov's type of spectrum by a strong electric field.
These investigations may play an essential role for the description of complex phenomena that appear
in dense astrophysical objects, as well as in the next generation intense laser-solid density plasma experiments.

\end{document}